\documentclass[a4paper,fleqn,final]{cas-dc}

\usepackage[T1]{fontenc}
\usepackage[utf8]{inputenc}
\usepackage[main=english]{babel}
\usepackage{textcomp}
\usepackage{hyperref}
\usepackage{amsmath,amssymb}
\usepackage{graphicx}
\usepackage{booktabs}
\usepackage{tabularx}
\usepackage{array}
\usepackage{calc}

\usepackage{threeparttable}
\usepackage{longtable}
\usepackage{adjustbox}
\usepackage[numbers,sort&compress]{natbib}
\usepackage{caption}
\usepackage{etoolbox}
\usepackage{placeins}
\usepackage{needspace}
\usepackage{stfloats}
\usepackage{xurl}
\usepackage{microtype}
\usepackage{lastpage}

\extrafloats{100}
\makeatletter
\g@addto@macro{\UrlBreaks}{\do\_\do\-\do\/\do\.\do\0\do\1\do\2\do\3\do\4\do\5\do\6\do\7\do\8\do\9}
\makeatother

\renewcommand{\ttfamily}{\rmfamily}
\AtBeginDocument{%
  \urlstyle{same}%
}

\usepackage{xcolor}
\definecolor{linknavy}{RGB}{0,70,127}
\hypersetup{
  colorlinks=true,
  linkcolor=black,
  citecolor=linknavy,
  urlcolor=linknavy,
  runcolor=linknavy
}

\AtBeginDocument{%
  \raggedbottom
  \setlength{\abovedisplayskip}{5pt plus 2pt minus 1pt}%
  \setlength{\belowdisplayskip}{5pt plus 2pt minus 1pt}%
  \setlength{\abovedisplayshortskip}{2pt plus 1pt}%
  \setlength{\belowdisplayshortskip}{4pt plus 1pt minus 1pt}%
}

\AtBeginEnvironment{table}{\let\sffamily\rmfamily}
\AtBeginEnvironment{figure}{\let\sffamily\rmfamily}
\AtBeginEnvironment{table*}{\let\sffamily\rmfamily}
\AtBeginEnvironment{figure*}{\let\sffamily\rmfamily}

\newcommand{\Rsq}{R^{2}}

\ExplSyntaxOn
\RenewDocumentCommand \firstname {}
  { \textcolor{black}{\seq_use:Nn \l_stm_au_seq { ~ }} }

\RenewDocumentCommand \emailauthor { m m }
   {
     \int_gincr:N \g_ead_int
     \seq_gput_right:Nn \g_stm_ead_seq
       {
         { \href{mailto:#1}{\rmfamily #1} }
         \parsename { #2 }
         \space(\eadauthor)
       }
     }

\cs_set:Npn \__first_footerline:
{
  \group_begin:
  \small
  \normalfont
  \ifnum\theblind>0\relax
  \else
  \__short_authors: :~
  \fi
  \itshape Preprint~submitted~to~Frontiers
  \group_end:
}

\RenewDocumentCommand \printorcid { } { }

\cs_set:Npn \__first_head:
{
  \parbox[t]{\textwidth}
  {
    \rule{\textwidth}{0pt}
  }
}

\cs_set:Npn \__cas_head:
{
  \parbox{\textwidth}
  {
    \rule{\textwidth}{0pt}
  }
}

\cs_set:Npn \__cas_foot:
{
  \parbox[t]{\textwidth}
  {
   \rule{\textwidth}{.2pt}\\
   \small
   \normalfont
   \__first_footerline:
   \hfill Page~\thepage {}~of~ \lastpage
  }
}
\ExplSyntaxOff

\makeatletter
\ps@cas
\makeatother

\begin{document}

\makeatletter
\def\bstctlcite{\@ifnextchar[{\@bstctlcite}{\@bstctlcite[@auxout]}}
\def\@bstctlcite[#1]#2{\@bsphack
  \@for\@citeb:=#2\do{%
    \edef\@citeb{\expandafter\@firstofone\@citeb}%
    \if@filesw\immediate\write\csname #1\endcsname{\string\citation{\@citeb}}\fi}%
  \@esphack}
\makeatother
\bstctlcite{IEEEbsTcontrol}

\shortauthors{Farajpoor and Narimani}
\shorttitle{GeoAI of snow after the Creek Fire}

\title[mode=title]{From Wildfire Severity to Snow Persistence: A Multisource GeoAI Study of the 2020 Creek Fire}

\author[1]{Parastoo Farajpoor}
\author[1]{Mohammadreza Narimani}
\cormark[1]
\ead{mnarimani@ucdavis.edu}

\affiliation[1]{organization={Department of Biological and Agricultural Engineering, University of California, Davis},city={Davis},state={CA},postcode={95616},country={USA}}

\cortext[1]{Corresponding author}

\begin{abstract}
Accurate prediction of post-fire snow conditions does not necessarily establish how wildfire changed those conditions. We developed an explainable geospatial artificial intelligence framework combining multisource Earth observations, meteorological information, and matched before--after comparisons to examine seasonal snow persistence following the 2020 Creek Fire in California's Sierra Nevada. Harmonized Landsat Sentinel-2 observations were used to estimate the fraction of clear-sky observations containing snow during October--July for eleven water years, 2016--2026, on a 500~m analysis grid. We matched 3,778 snow-zone cells inside the fire perimeter to comparable unburned controls using terrain and pre-fire snow conditions; absolute standardized mean differences after matching were no greater than 0.047. HLS persistence agreed closely with MODIS, with a mean annual spatial correlation of 0.947. The landscape-average before--after control--impact contrast was $+0.0017$, with a year-level 95\% confidence interval of $-0.023$ to $+0.027$. A stronger response emerged in the highest burn-severity class, where observed persistence increased by 0.026 relative to matched controls, equivalent to 2.6 percentage points. Under 5~km spatial cross-validation, XGBoost predicted raw post-fire persistence with an out-of-fold coefficient of determination of 0.811, whereas predictive skill for the fire-adjusted anomaly reached 0.046. Elevation and temperature together accounted for 65.7\% of mean absolute model attribution. These findings distinguish a severity-associated optical snow response from the terrain--climate relationships that dominate predictive skill. Combining matched comparisons with explainable GeoAI provides a practical framework for forest monitoring that separates accurate environmental mapping from inference about disturbance effects.
\end{abstract}


\maketitle

\section{Introduction}\label{introduction}
Mountain snow links forest condition to the timing and availability of water across the western United States. Warming is changing the amount, duration, and seasonal distribution of snow, while wildfire increasingly overlaps landscapes that historically retained snow through winter and spring \citep{SiirilaWoodburn2021_2100219y,Kampf2022_00333119,Koshkin2022_22971271}. Where these changes coincide, forest disturbance can alter both the snowpack and the observations used to monitor it. Understanding that distinction is important for interpreting satellite records and for translating artificial intelligence into useful assessments of post-fire environmental change.

The snow response to wildfire is not governed by a single mechanism. Canopy removal can reduce interception and increase the proportion of snowfall reaching the ground, but it also changes shading, longwave radiation, wind exposure, and snow redistribution \citep{Varhola2010_01008009,Lundquist2013_rcr20504}. Charred material deposited on snow can lower albedo and increase absorbed solar energy, accelerating ablation \citep{Gleason2013_grl50896,Gleason2019_1909935y}. The resulting balance depends on meteorological conditions, terrain, and the degree of forest disturbance \citep{Maxwell2019_ab5de8,Moeser2020_WR027071}. Greater snow accumulation and faster melt can therefore occur within the same burned landscape.

The quantity being measured is equally consequential. Satellite snow cover, snow disappearance date, snow depth, and snow water equivalent describe related but different aspects of the seasonal snowpack. Following the Moonlight Fire in the Sierra Nevada, \citet{Micheletty2014_46012014} documented increased fractional snow cover using MODIS-derived observations. More recent work has examined changes in snow disappearance under different climatic conditions and used airborne lidar with machine learning to estimate seasonally varying changes in snow depth across Sierra Nevada basins \citep{Koshkin2025_vadt9866,Koshkin2026_34672026}. These studies show why an apparent increase in optical snow presence need not imply greater water storage or later disappearance.

Multisource Earth observation now makes it possible to evaluate such differences over long records. Harmonized Landsat Sentinel-2 imagery provides a common framework for combining medium-resolution optical observations, while reanalysis and elevation products describe environmental conditions that influence snow \citep{Ju2025_HLS2,MunozSabater2021_essd4349}. Geospatial artificial intelligence, or GeoAI, offers tools for integrating these observations, representing nonlinear relationships, and examining how predictions vary geographically \citep{Janowicz2020_91684500,Mai2025_GeoAI}. The scientific value of this integration, however, depends on the target, comparison design, and evaluation strategy rather than on model complexity alone.

Recent applications illustrate the diversity of targets addressed by optical GeoAI. Active-fire segmentation identifies burning pixels, whereas post-fire multisource analysis characterizes subsequent landscape responses \citep{Mitra2026_ActiveFire,Farajpoor2026_Kincade}. Satellite time-series studies in agricultural systems similarly emphasize the importance of seasonal alignment and environmental context when interpreting spectral changes \citep{Narimani2025_Broomrape,Narimani2026c_Yield}. Across these applications, a model can predict an environmental state accurately without isolating the effect of a particular disturbance. Spatially structured evaluation is therefore essential, especially when nearby observations share terrain, climate, and acquisition conditions \citep{Roberts2017_cog02881,Ploton2020_2018321y}.

The 2020 Creek Fire provides a focused setting in which to examine this problem. The fire affected a large, topographically varied landscape in the central Sierra Nevada, including montane forests with substantial pre-existing tree mortality \citep{Stephens2022_22120258}. Five pre-fire and six post-fire snow seasons allow the post-fire record to be interpreted against both local baseline conditions and contemporaneous unburned controls. Rather than treating a high predictive score as evidence of a fire effect, we evaluate two targets separately: observed seasonal snow persistence and its matched, pre-fire-adjusted anomaly.

\begin{figure*}[!t]
\centering
\includegraphics[width=\textwidth]{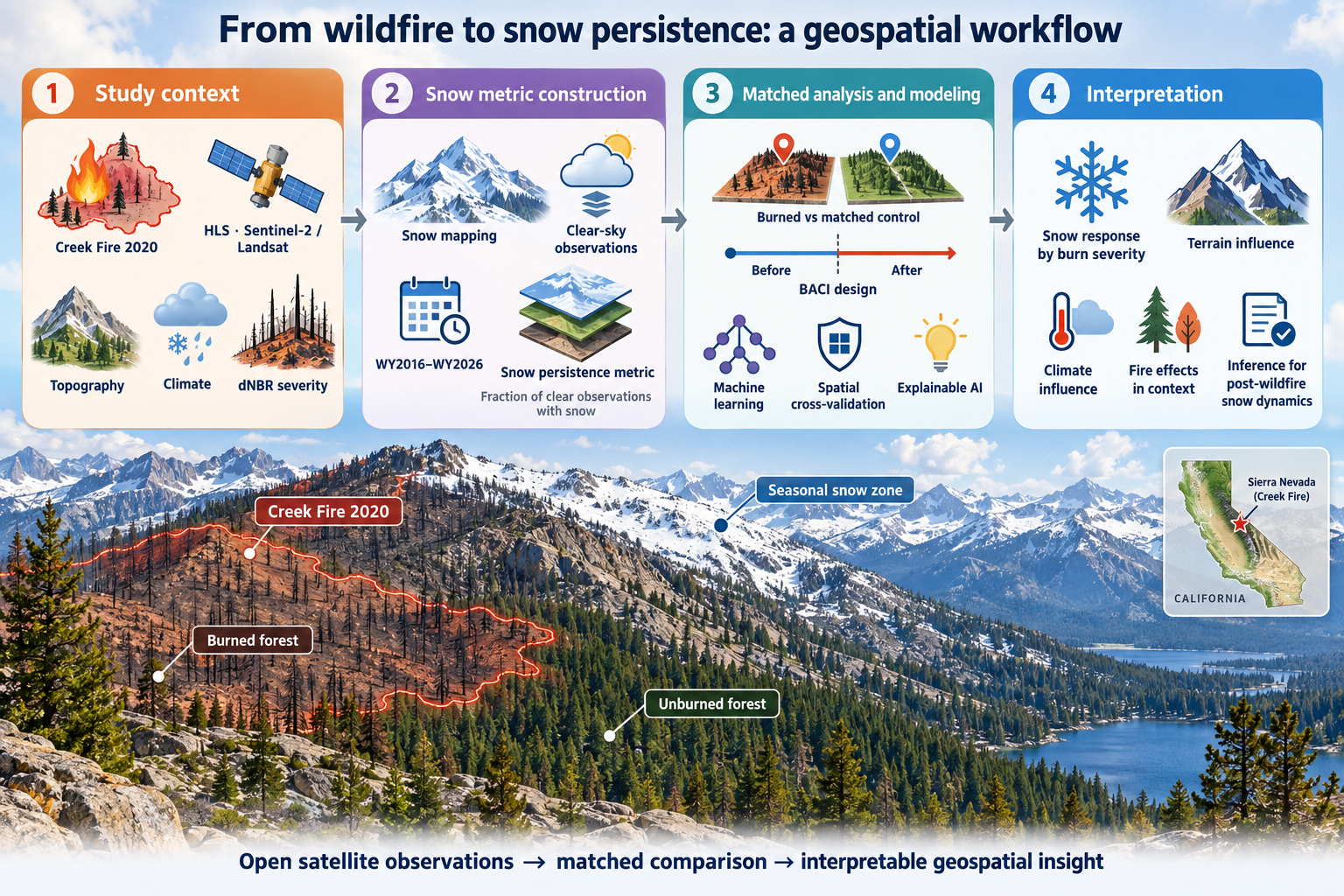}
\caption{Multisource GeoAI framework for examining post-fire snow persistence. Satellite observations, terrain, meteorological information, and spectral burn severity support snow-metric construction, matched before--after comparisons, spatially evaluated machine learning, and interpretation. The lower landscape is a conceptual illustration; its terrain and fire outline are symbolic rather than a georeferenced reconstruction of the study area.}
\label{fig:workflow}
\end{figure*}

This study addresses three questions. First, did observed snow persistence inside the Creek Fire perimeter change relative to comparable unburned locations? Second, did the response vary among spectral burn-severity classes? Third, could terrain, meteorology, and fire descriptors predict the adjusted response as effectively as they predicted snow persistence itself? The framework combines multisource observations, matching, before--after control--impact analysis, spatially blocked machine learning, and model attribution (Figure~\ref{fig:workflow}). Its contribution is an integrated assessment of what the satellite record shows, what the matched comparison supports, and what GeoAI can predict.

\section{Materials and methods}\label{materials-and-methods}

\subsection{Study area and observation period}\label{study-area}
The study examined the 2020 Creek Fire in California's Sierra Nevada, approximately centered at $37.33^{\circ}$\,N, $119.28^{\circ}$\,W. The fire-perimeter record was identified by the Monitoring Trends in Burn Severity event identifier CA3720111927220200905. MTBS provides a consistent framework for identifying and documenting large fire events, although the spectral severity measurements used here were derived separately from Sentinel-2 imagery \citep{Eidenshink2007_y0301003}.

\begin{figure*}[!t]
\centering
\includegraphics[width=\textwidth]{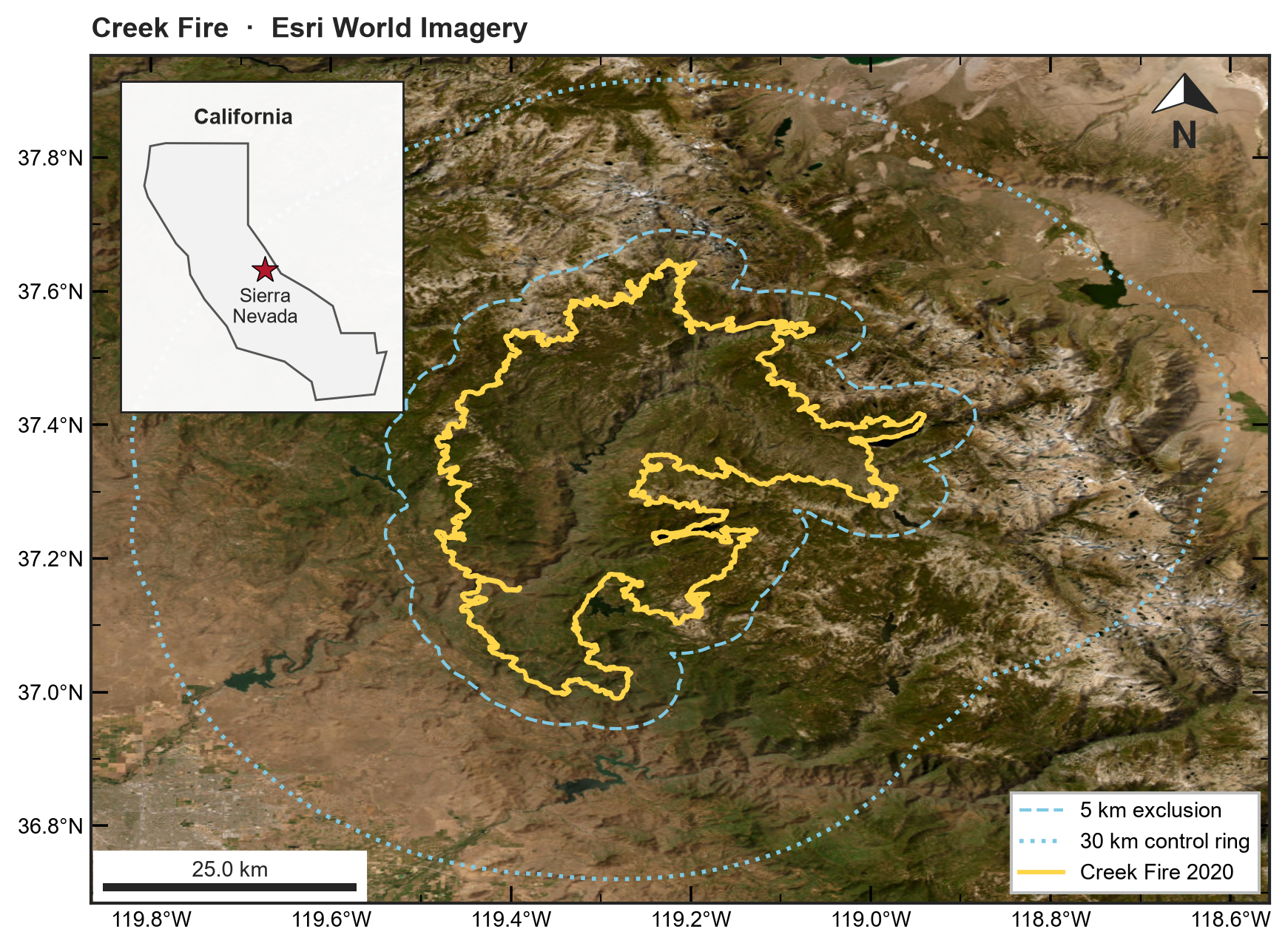}
\caption{Creek Fire study area and candidate-control region. The yellow outline marks the 2020 fire perimeter. The dashed and dotted boundaries indicate the 5~km inner exclusion and 30~km outer boundary, respectively. The inset locates the study within California's Sierra Nevada. Basemap: Esri World Imagery.}
\label{fig:setting}
\end{figure*}

The analysis domain extended from $119.83^{\circ}$ to $118.59^{\circ}$\,W and from $36.64^{\circ}$ to $38.00^{\circ}$\,N. Candidate controls were selected from an annulus extending 5--30~km beyond the perimeter (Figure~\ref{fig:setting}). The inner separation reduced immediate boundary mixing, while the outer boundary retained a regional comparison pool. Geographic proximity was not treated as sufficient evidence of comparability; terrain and pre-fire snow conditions were subsequently used for matching.

The analytical record covered water years WY2016--WY2026. Each snow-observation season ran from 1 October of the preceding calendar year through 31 July of the named water year. WY2016--WY2020 constituted the pre-fire period, and WY2021--WY2026 constituted the post-fire period. Thus, the WY2020 snow-observation season ended before the September 2020 fire.

All statistical analyses used a 500~m grid in WGS~84/UTM zone 11N, EPSG:32611. The grid contained 36,456 cells before snow-zone filtering: 6,177 inside the perimeter and 30,279 in the candidate-control region. Retaining cells with mean pre-fire persistence greater than 0.05 produced 21,284 eligible cells, comprising 3,778 perimeter-internal cells and 17,506 candidate controls. Perimeter-internal cells are termed treatment cells below; this designation identifies their role in the comparison rather than implying that every cell experienced substantial combustion.

\subsection{Satellite observations and seasonal snow persistence}\label{snow-persistence}
Snow observations were derived from NASA HLS version 2 surface reflectance using the Earth Engine collections \texttt{NASA/HLS/HLSL30/v002} and \texttt{NASA/HLS/HLSS30/v002}. HLS combines Landsat and Sentinel-2 observations on a common 30~m grid using radiometric and geometric harmonization procedures \citep{Claverie2018_01809002,Ju2025_HLS2}. Pixels flagged for cirrus, cloud, cloud adjacency, cloud shadow, or water were excluded.

Snow classification used the Normalized Difference Snow Index:
\begin{equation}
\mathrm{NDSI}_{xk}
=
\frac{\rho_{\mathrm{green},xk}-\rho_{\mathrm{SWIR1},xk}}
{\rho_{\mathrm{green},xk}+\rho_{\mathrm{SWIR1},xk}}
\label{eq:ndsi}
\end{equation}
where $x$ denotes a location, $k$ an acquisition, and $\rho$ surface reflectance. Equation~\eqref{eq:ndsi} used bands B3 and B6 for HLSL30 and B3 and B11 for HLSS30. A valid observation was classified as snow when NDSI exceeded 0.40 and green reflectance exceeded 0.10. This spectral contrast exploits the difference between snow reflectance in visible and shortwave-infrared wavelengths \citep{Dozier1989_89901016}.

For a location and season, observed snow persistence was calculated as
\begin{equation}
\begin{aligned}
P_{xt}
&=
\frac{\sum_{k\in \mathcal{K}_t} q_{xk}s_{xk}}
{\sum_{k\in \mathcal{K}_t} q_{xk}},\\[2pt]
s_{xk}
&=
\mathbf{1}
\left(
\mathrm{NDSI}_{xk}>0.40
\;\land\;
\rho_{\mathrm{green},xk}>0.10
\right)
\end{aligned}
\label{eq:persist}
\end{equation}
where $\mathcal{K}_t$ contains acquisitions in the October--July window, $q_{xk}$ indicates a valid clear-sky observation, and $s_{xk}$ is the binary snow flag. Seasonal products were exported on the 500~m analytical grid.

Equation~\eqref{eq:persist} defines an observation-frequency metric between zero and one. It measures the fraction of valid acquisitions containing detectable snow, not the fraction of calendar days with snow. This distinction separates the present outcome from daily snow duration and snow disappearance date. Snow persistence is useful for describing seasonal snow occurrence \citep{Hammond2018_joc5674}, but its optical implementation remains sensitive to canopy visibility, acquisition timing, and unresolved mixtures of snow and other surfaces \citep{Stillinger2023_75672023}.

A comparison record was generated from MODIS/Terra MOD10A1 Collection 6.1 \citep{HallRiggs2021_MOD10A1}. MODIS persistence used the same seasonal window, a snow threshold of \texttt{NDSI\_Snow\_Cover} greater than 40, and exclusion of values at or above 200 representing non-valid snow-cover categories. Annual spatial Pearson and Spearman correlations and domain-mean persistence were used to evaluate consistency between the two optical records. These comparisons assessed cross-sensor agreement rather than accuracy against ground observations.

\subsection{Terrain, spectral disturbance, and meteorology}\label{terrain-severity}
Elevation was obtained from Copernicus GLO-30 and averaged to the analytical grid. Copernicus DEM is a digital surface model, so its elevation surface can include vegetation and infrastructure as well as terrain. Slope and aspect were derived using finite differences. Aspect was represented as northness, $\cos(\mathrm{aspect})$, and eastness, $\sin(\mathrm{aspect})$, avoiding an artificial discontinuity between $0^{\circ}$ and $360^{\circ}$. A dimensionless exposure proxy, $\cos(\mathrm{slope})\max[0,-\mathrm{northness}]$, summarized slope orientation; it was not interpreted as a physically modeled radiation flux.

Spectral disturbance was calculated from Sentinel-2 Level-2A imagery (\texttt{COPERNICUS/\allowbreak S2\_SR\_HARMONIZED}). Pre-fire and post-fire composites were the medians of scenes with less than 40\% reported cloud cover during 1 June--15 August 2020 and 1 October--30 November 2020, respectively. The Normalized Burn Ratio and its difference were
\begin{equation}
\begin{aligned}
\mathrm{NBR}
&=
\frac{\rho_{\mathrm{NIR}}-\rho_{\mathrm{SWIR2}}}
{\rho_{\mathrm{NIR}}+\rho_{\mathrm{SWIR2}}},\\[2pt]
\mathrm{dNBR}
&=
\mathrm{NBR}_{\mathrm{pre}}
-
\mathrm{NBR}_{\mathrm{post}}
\end{aligned}
\label{eq:dnbr}
\end{equation}
using Sentinel-2 bands B8 and B12. Equation~\eqref{eq:dnbr} provides a spectral disturbance indicator whose interpretation depends on vegetation, background conditions, and spatial support \citep{Miller2007_00612006}.

The study-defined dNBR classes were less than 0.10, 0.10--0.27, 0.27--0.44, and at least 0.44, labeled unburned, low, moderate, and high, respectively. These labels describe the analytical spectral classes and are not field-calibrated MTBS severity assignments. In particular, ``unburned'' inside the perimeter denotes low spectral change rather than an independently verified absence of fire exposure.

\begin{figure*}[!t]
\centering
\includegraphics[width=\textwidth]{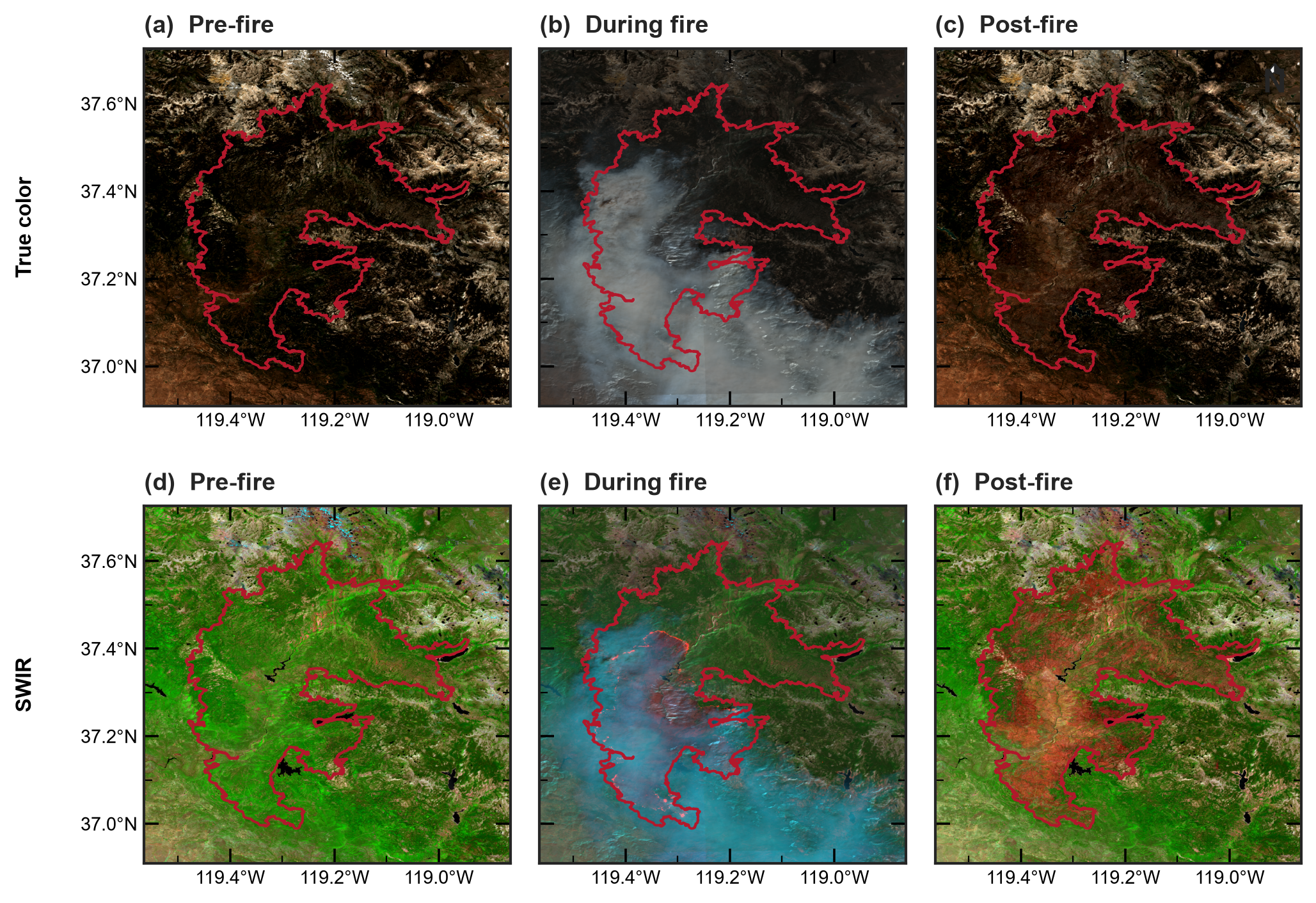}
\caption{Sentinel-2 observations of the Creek Fire landscape. Panels A--C show true-color composites using B4--B3--B2; panels D--F show shortwave-infrared composites using B12--B8--B4. The displayed periods are July--August 2020, 8 September 2020, and July--August 2021. These images illustrate landscape conditions and are distinct from the pre-fire and post-fire composite windows used for the dNBR calculation.}
\label{fig:fire}
\end{figure*}

The true-color and shortwave-infrared composites in Figure~\ref{fig:fire} provide visual context for the event. Their display dates differ from the composite windows used to calculate dNBR: July--August 2020 before the fire, 8 September 2020 during the fire, and July--August 2021 afterward.

Meteorological predictors were obtained from ERA5-Land \citep{MunozSabater2021_essd4349}. Domain-mean summaries of 2~m air temperature, precipitation, downward shortwave radiation, and positive degree-days were assigned to each water year. Every cell within a given year received the same meteorological values. Consequently, these predictors represented interannual meteorological variation rather than local climatic differences within the fire perimeter.

\subsection{Matching and before--after comparison}\label{matching}
Matching used only pre-fire snow conditions and terrain variables. Each treatment cell was paired with its nearest eligible control in the standardized six-dimensional space of elevation, slope, northness, eastness, exposure proxy, and mean pre-fire persistence. For treatment cell $i$ and candidate control $j$, the matching distance was
\begin{equation}
\begin{aligned}
d(i,j)
&=
\left[
\sum_{\ell=1}^{6}
\left(z_{i\ell}-z_{j\ell}\right)^2
\right]^{1/2},\\[2pt]
m(i)
&=
\underset{j}{\operatorname{argmin}}\;d(i,j)
\end{aligned}
\label{eq:match}
\end{equation}
where $z$ denotes standardized covariates and $m(i)$ the selected control.

Matching was one-to-one with replacement, with a caliper of $0.5\sqrt{6}$, approximately 1.22. All 3,778 treatment cells were retained; the maximum selected distance was 1.11. The matched sample contained 2,157 distinct control cells, some of which were used for more than one treatment cell. Table~\ref{tab:sample} summarizes the eligible and matched samples. Standardized mean differences were used to evaluate balance, following established approaches to environmental matching \citep{Schleicher2020_obi13448}.

\begin{table*}[!t]
\centering
\caption{Snow-zone cells before and after matching.}
\label{tab:sample}
\footnotesize
\resizebox{\textwidth}{!}{%
\begin{tabular}{lccccc}
\toprule
Group & Sample size & Mean elevation, m (range) & Mean pre-fire persistence & Mean slope, $^{\circ}$ & Mean northness \\
\midrule
Treatment cells inside the perimeter & 3,778 & 2,142 (1,351--2,957) & 0.246 & 8.7 & $-0.20$ \\
Eligible control pool & 17,506 & 2,776 (184--4,034) & 0.467 & --- & --- \\
Matched controls & 3,778 & 2,154 (1,314--3,021) & 0.245 & 8.6 & $-0.20$ \\
\bottomrule
\end{tabular}}\\[2pt]
\parbox{\textwidth}{\footnotesize Eligibility required mean pre-fire persistence greater than 0.05. Matched-control summaries reflect 3,778 pair entries mapped to 2,157 distinct control cells, including repeated control use. Dashes indicate quantities not reported for the full control pool.}
\end{table*}

Eligibility required mean pre-fire persistence greater than 0.05. Matched-control summaries reflect the matched sample, including repeated control use. Dashes in Table~\ref{tab:sample} indicate quantities not reported for the full control pool.

For pair $i$ and water year $t$, the contemporaneous treatment--control difference was $D_{it}$. The cell-level before--after control--impact contrast and its sample mean were
\begin{equation}
\begin{aligned}
D_{it}
&=
P^{T}_{it}
-
P^{C}_{m(i)t},\\[2pt]
\delta_i
&=
\frac{1}{6}\sum_{t=2021}^{2026}D_{it}
-
\frac{1}{5}\sum_{t=2016}^{2020}D_{it},\\[2pt]
\widehat{\delta}
&=
\frac{1}{N}\sum_{i=1}^{N}\delta_i
\end{aligned}
\label{eq:baci}
\end{equation}
Here, $T$ and $C$ identify treatment and control cells. Positive values in Equation~\eqref{eq:baci} indicate a greater post-fire increase, or smaller decrease, in observed persistence at treatment cells relative to their controls.

The matched BACI design reduces confounding from baseline differences and shared annual conditions, but its interpretation remains conditional on the comparison design \citep{Butsic2017_01701005}. Matching balance does not by itself establish that all relevant environmental differences have been removed.

Primary uncertainty was evaluated at the water-year level. Let $\overline{D}_t$ be the mean matched difference in year $t$, with pre-fire and post-fire sample variances $s_{\mathrm{pre}}^2$ and $s_{\mathrm{post}}^2$. The Welch interval was
\begin{equation}
\widehat{\delta}
\;\pm\;
t_{0.975,\nu}
\left(
\frac{s_{\mathrm{post}}^{2}}{6}
+
\frac{s_{\mathrm{pre}}^{2}}{5}
\right)^{1/2}
\label{eq:welch}
\end{equation}
where $\nu$ is the Welch--Satterthwaite degrees of freedom. Equation~\eqref{eq:welch} treats the five pre-fire and six post-fire annual summaries as the temporal replicate units. Cell-level intervals are also reported to describe uncertainty conditional on the observed years. The distinction matters because numerous spatial observations within one fire do not provide replication across independent fire events \citep{Underwood1994_1942110}.

Class-wise contrasts were calculated for the four dNBR categories. The relationship between continuous dNBR and $\delta_i$ was summarized using Spearman correlation. Elevation-tercile and north-facing versus south-facing comparisons were treated as exploratory.

\subsection{Spatial prediction and model attribution}\label{prediction}
Machine-learning models were evaluated on two post-fire targets. The first was the fire-adjusted anomaly,
\begin{equation}
A_{it}
=
D_{it}
-
\frac{1}{5}\sum_{u=2016}^{2020}D_{iu},
\qquad t\geq 2021
\label{eq:anomaly}
\end{equation}
which subtracts each pair's pre-fire mean difference from its contemporaneous treatment--control contrast. The second was raw treatment-cell persistence, $P^{T}_{it}$. Equation~\eqref{eq:anomaly} defines an adjusted observational response; it does not convert that response into a directly measured causal fire effect.

The modeling dataset contained 22,665 cell-years. Elastic Net, Random Forest, and XGBoost represented regularized linear, bagged-tree, and boosted-tree approaches, respectively \citep{ZouHastie2005_ElasticNet,Breiman2001_33404324,Chen2016_22939785}. Elastic Net used a regularization parameter of 0.05 and an L1 ratio of 0.5. Random Forest used 250 trees and a minimum leaf size of five. XGBoost used 250 trees, maximum depth five, a learning rate of 0.05, and row and column subsampling fractions of 0.8.

Models were evaluated using four reported feature configurations: terrain alone; terrain plus meteorology; the addition of dNBR and years since fire; and the full configuration. The terrain-only configuration contained elevation, slope, and northness. Continuous inputs were standardized using training-fold information only.

Five-fold GroupKFold cross-validation used 5~km spatial blocks. Observations associated with a cell remained in the same fold. Out-of-fold $\Rsq$, root-mean-square error (RMSE), and mean absolute error (MAE) summarized predictive performance. Spatial grouping reduces the direct overlap of nearby training and test observations compared with random cell-level splitting \citep{Roberts2017_cog02881,Ploton2020_2018321y}.

Every post-fire year was represented in both training and test regions. The evaluation therefore addressed prediction at withheld spatial locations during the observed years, rather than forecasting an unseen year or transferring the model to a different fire. This distinction was particularly relevant for the domain-mean meteorological predictors, which could also distinguish years.

Model attribution used TreeExplainer SHAP values for 3,000 rows of the fitted XGBoost persistence model \citep{Lundberg2020_01901389}. Global feature importance and the attribution share of a feature family were
\begin{equation}
\begin{aligned}
I_j
&=
\frac{1}{M}\sum_{r=1}^{M}|\phi_{rj}|,\\[2pt]
S_{\mathcal F}
&=
100\,
\frac{\sum_{j\in\mathcal F}I_j}
{\sum_{j=1}^{p}I_j}
\end{aligned}
\label{eq:shap}
\end{equation}
where $\phi_{rj}$ is the SHAP value for feature $j$ and observation $r$, $M=3{,}000$, and $\mathcal F$ denotes terrain, climate, or fire descriptors. Equation~\eqref{eq:shap} summarizes attribution magnitude. These shares are not percentages of physical causation or independently explained variance.

\subsection{Software and reproducibility}\label{software}
Satellite processing used Google Earth Engine. Subsequent analysis used Python with GDAL, GeoPandas, NumPy, pandas, SciPy, scikit-learn, XGBoost, and SHAP. A random seed of 42 was used for stochastic analysis steps. The reproducibility release will include the derived analytical panel, matching information, model configurations, out-of-fold predictions, figure-source outputs, and environment specifications.

\section{Results}\label{results}

\subsection{Terrain and pre-fire snow conditions define the comparison}\label{results-terrain}
Before matching, the candidate-control pool occupied substantially higher and snowier terrain than the treatment sample. Mean elevation was 2,776~m among eligible controls and 2,142~m inside the perimeter; corresponding pre-fire persistence means were 0.467 and 0.246 (Table~\ref{tab:sample}). Directly comparing these groups would therefore have combined fire exposure with a strong pre-existing environmental contrast.

Matching substantially reduced these differences. The absolute standardized mean difference decreased from 1.593 to 0.047 for elevation and from 1.189 to 0.013 for pre-fire persistence. All six matched covariates had absolute standardized mean differences no greater than 0.047 (Table~\ref{tab:balance}). The matched controls consequently represented a much closer baseline comparison than the full regional control pool.

\begin{table*}[!t]
\centering
\caption{Standardized mean differences before and after matching.}
\label{tab:balance}
\footnotesize
\begin{tabularx}{\textwidth}{@{}Xcc@{}}
\toprule
Covariate & Before matching & After matching \\
\midrule
Elevation & $-1.593$ & $-0.047$ \\
Slope & $-0.246$ & $+0.017$ \\
Northness & $-0.312$ & $+0.004$ \\
Eastness & $-0.061$ & $+0.002$ \\
Exposure proxy & $+0.289$ & $-0.002$ \\
Pre-fire persistence & $-1.189$ & $+0.013$ \\
\bottomrule
\end{tabularx}\\[2pt]
\parbox{\textwidth}{\footnotesize Signs indicate treatment minus control. Balance concerns the measured covariates and does not establish equivalence on unmeasured characteristics.}
\end{table*}

\begin{figure*}[!t]
\centering
\includegraphics[width=\textwidth]{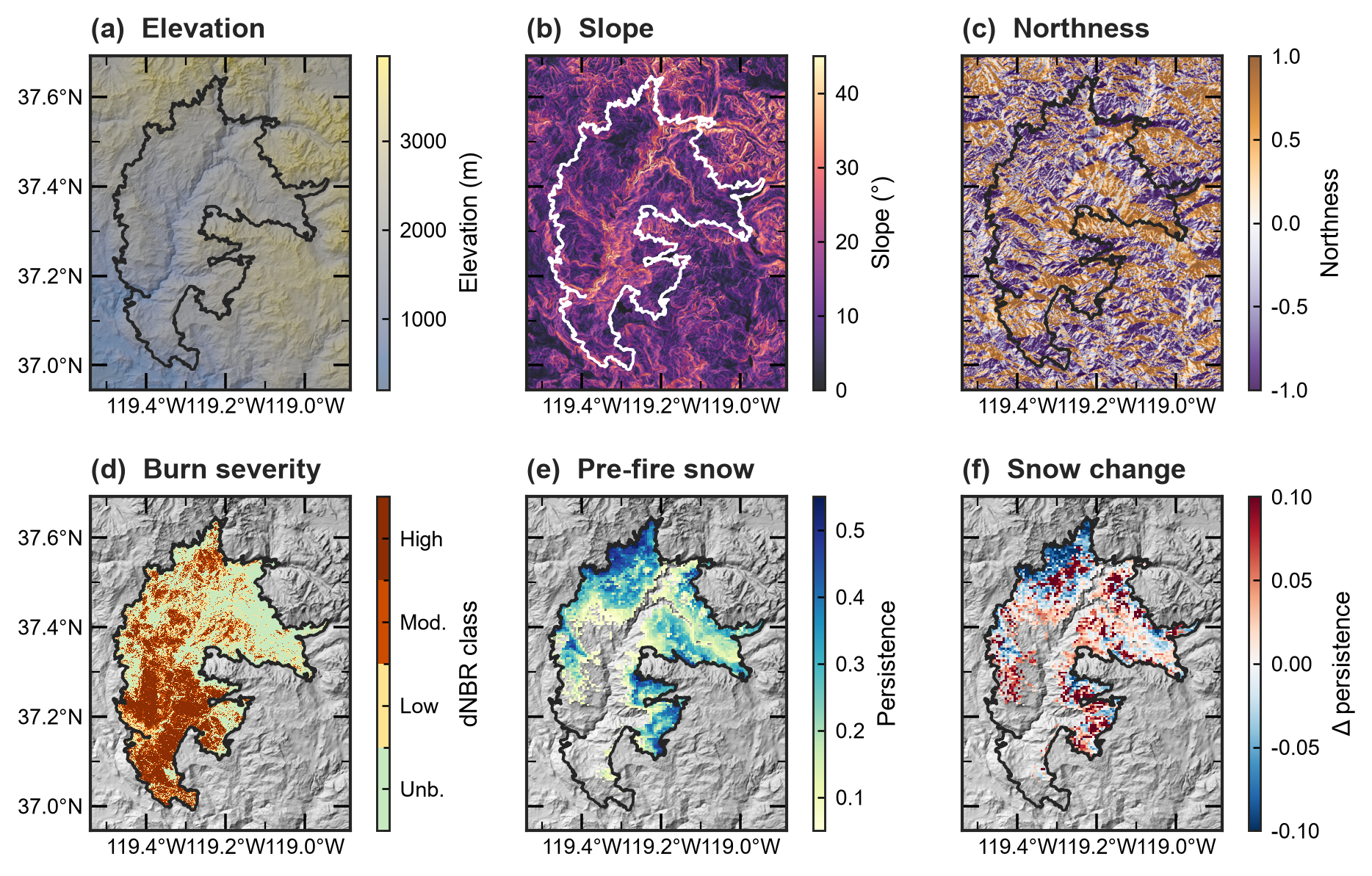}
\caption{Terrain, spectral disturbance, and observed snow persistence. Panels show elevation (A), slope (B), northness (C), study-defined Sentinel-2 dNBR classes (D), mean pre-fire HLS persistence (E), and raw post-fire minus pre-fire persistence (F). Panel F is not control-adjusted. ``Unb.'' denotes the low-change dNBR class inside the perimeter, not the external control sample.}
\label{fig:terrain}
\end{figure*}

\begin{figure*}[!t]
\centering
\includegraphics[width=\textwidth]{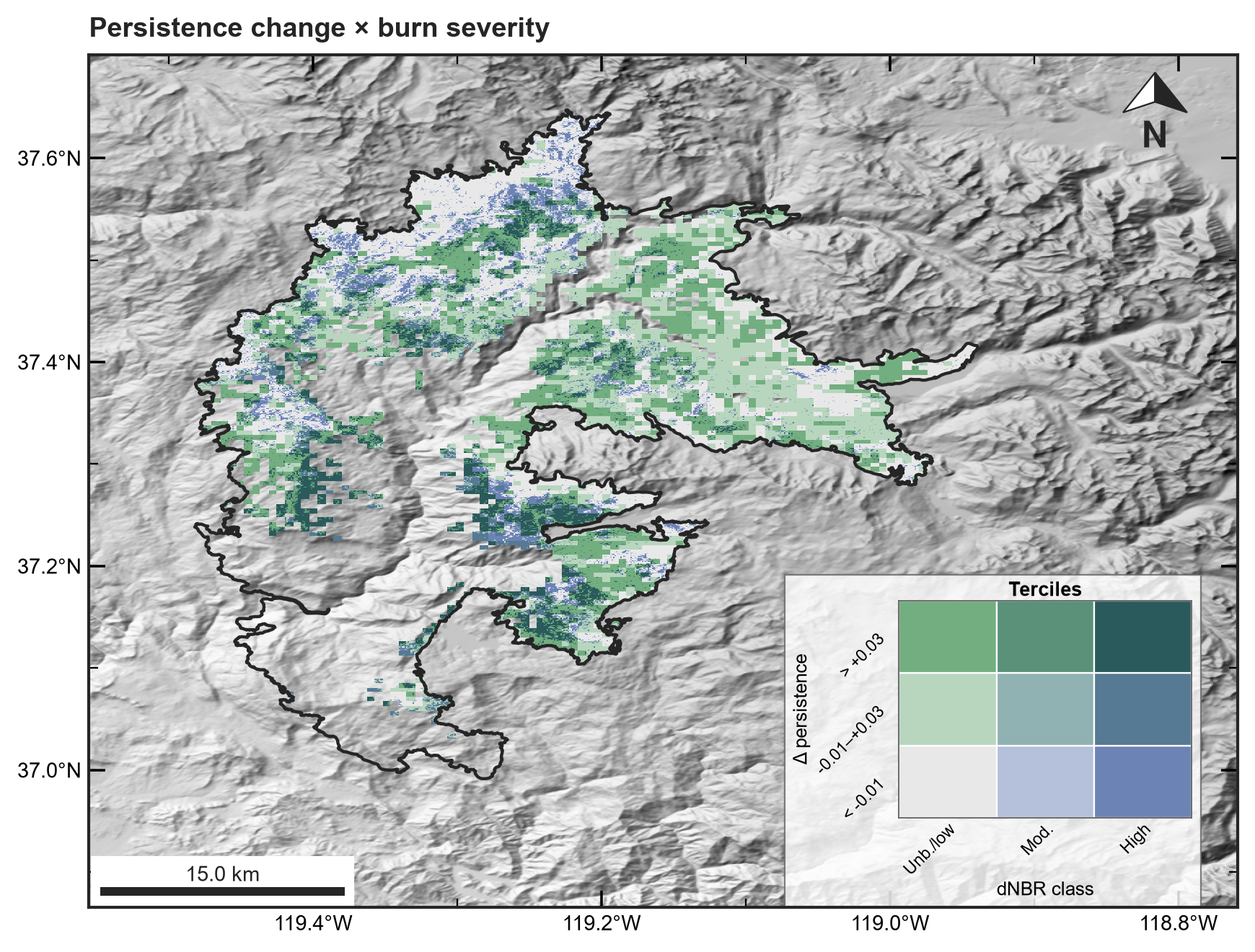}
\caption{Spatial correspondence between spectral burn severity and raw persistence change. Colors combine dNBR classes with post-fire minus pre-fire HLS persistence for eligible cells inside the perimeter. The displayed change categories are less than 0.01, 0.01--0.03, and greater than 0.03. This descriptive map does not subtract matched-control changes.}
\label{fig:bivariate}
\end{figure*}

The spatial layers show why this adjustment was necessary. Pre-fire persistence followed the elevation structure of the landscape, whereas dNBR and post--pre persistence change formed more heterogeneous patterns (Figure~\ref{fig:terrain}). The bivariate map further showed that spectral disturbance and raw snow-persistence change did not vary uniformly across the perimeter (Figure~\ref{fig:bivariate}). These maps describe unadjusted spatial patterns; the matched contrasts provide the subsequent test of change relative to controls.

\subsection{The landscape-average contrast is small relative to annual variation}\label{results-mean}
The mean matched treatment--control difference was $+0.0017$ before the fire and $+0.0034$ afterward, yielding an overall BACI contrast of $+0.0017$ (Table~\ref{tab:sev}). The cell-level interval was positive, 0.0003--0.0032, but the year-level interval extended from $-0.023$ to $+0.027$, with $p=0.88$. At the scale of annual replication, the average change was therefore not distinguishable from interannual variation.

Both treatment and matched-control cells had higher mean persistence in the post-fire period. Treatment means increased from 0.246 to 0.261, while matched-control means increased from 0.245 to 0.257. Their similar movement explains why the adjusted contrast was much smaller than the raw before--after change. The median cell-level BACI contrast was $-0.003$, and 47.2\% of cells had positive contrasts.

Annual matched differences also varied in sign. Post-fire annual means ranged from $-0.013$ to $+0.021$, within the range observed before the fire. The overall result thus provides no resolved landscape-average increase or decrease; it does not demonstrate equivalence between burned and unburned conditions.

\subsection{Higher spectral severity is associated with increased observed persistence}\label{results-severity}
\begin{figure*}[!t]
\centering
\includegraphics[width=\textwidth]{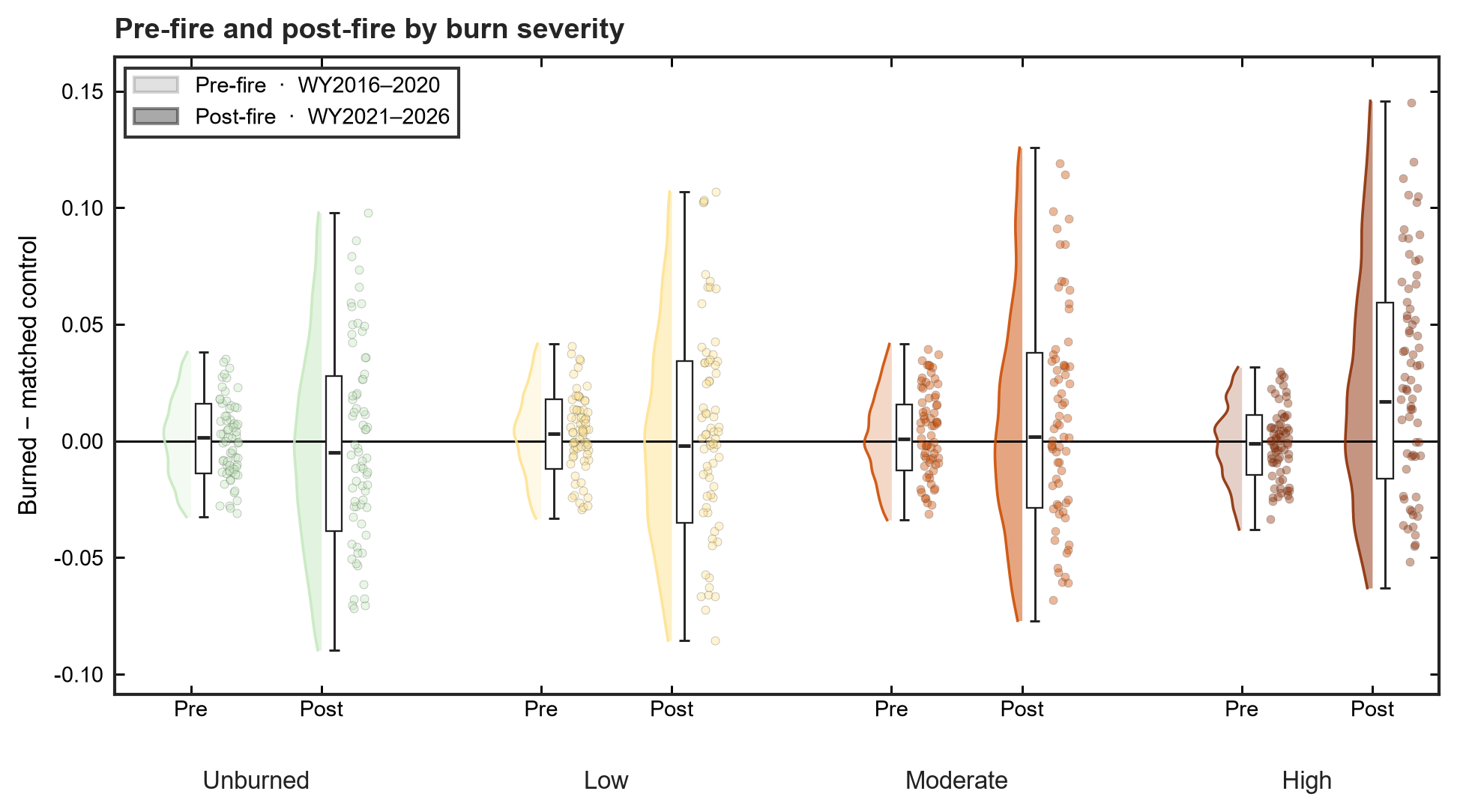}
\caption{Matched treatment--control differences before and after the Creek Fire. For each dNBR class, the left distribution represents WY2016--WY2020 and the right distribution represents WY2021--WY2026. Raincloud displays combine a density estimate, box plot, and subsample of cell values. The plotted distributions omit values outside the 5th--95th percentiles for display. ``Unburned'' refers to the perimeter-internal dNBR $< 0.10$ class.}
\label{fig:raincloud}
\end{figure*}

The highest dNBR class showed a different response from the perimeter-wide mean. Its 448 cells shifted from a mean matched difference of $-0.0014$ before the fire to $+0.0246$ afterward. The corresponding BACI contrast was $+0.0260$, or 2.60 percentage points of observed persistence, with a year-level 95\% interval of $+0.002$ to $+0.050$. Approximately 66\% of cells in this class had positive contrasts (Table~\ref{tab:sev}; Figure~\ref{fig:raincloud}).

\begin{table*}[!t]
\centering
\caption{Matched BACI contrasts in observed HLS snow persistence by dNBR class.}
\label{tab:sev}
\footnotesize
\resizebox{\textwidth}{!}{%
\begin{tabular}{lrrrrlll}
\toprule
Class & $n$ & Pre-fire $D$ & Post-fire $D$ & BACI contrast & Cell-level 95\% interval & Year-level 95\% interval & Positive cell contrasts \\
\midrule
All matched cells & 3,778 & $+0.0017$ & $+0.0034$ & $+0.0017$ & $+0.0003$ to $+0.0032$ & $-0.023$ to $+0.027$ & 47\% \\
Internal low-change, dNBR $< 0.10$ & 1,788 & $+0.0016$ & $-0.0026$ & $-0.0042$ & $-0.0068$ to $-0.0015$ & $-0.028$ to $+0.019$ & 44\% \\
Low, 0.10--0.27 & 977 & $+0.0031$ & $+0.0014$ & $-0.0018$ & $-0.0053$ to $+0.0018$ & $-0.030$ to $+0.026$ & 44\% \\
Moderate, 0.27--0.44 & 565 & $+0.0021$ & $+0.0092$ & $+0.0070$ & $+0.0019$ to $+0.0122$ & $-0.023$ to $+0.037$ & 49\% \\
High, $\ge 0.44$ & 448 & $-0.0014$ & $+0.0246$ & $+0.0260$ & $+0.0200$ to $+0.0321$ & $+0.002$ to $+0.050$ & 66\% \\
\bottomrule
\end{tabular}}\\[2pt]
\parbox{\textwidth}{\footnotesize $D$ is treatment minus matched-control persistence. Year-level intervals use five pre-fire and six post-fire annual means. Cell-level intervals condition on the observed years and are secondary to the year-level inference. Persistence and its contrasts are dimensionless fractions. Values are rounded independently.}
\end{table*}

The moderate class had a smaller positive contrast of $+0.0070$, while low-severity and perimeter-internal low-change cells had contrasts of $-0.0018$ and $-0.0042$, respectively. Their year-level intervals included zero. Although the class means increased in order with severity, the continuous association between dNBR and cell-level BACI was weak, with Spearman $\rho=0.11$. Severity therefore distinguished a class-level pattern more clearly than a strong, uniformly graded cell-level relationship.

Exploratory elevation-tercile contrasts were $+0.002$, $+0.004$, and $-0.001$, without a monotonic elevation pattern. North-facing cells had a mean contrast of $+0.010$ compared with $-0.004$ for south-facing cells. These aspect summaries were not adjusted for differences in severity composition and are interpreted as descriptive contrasts.

\subsection{HLS and MODIS capture consistent spatial and annual snow patterns}\label{results-modis}
\begin{figure*}[!t]
\centering
\includegraphics[width=\textwidth]{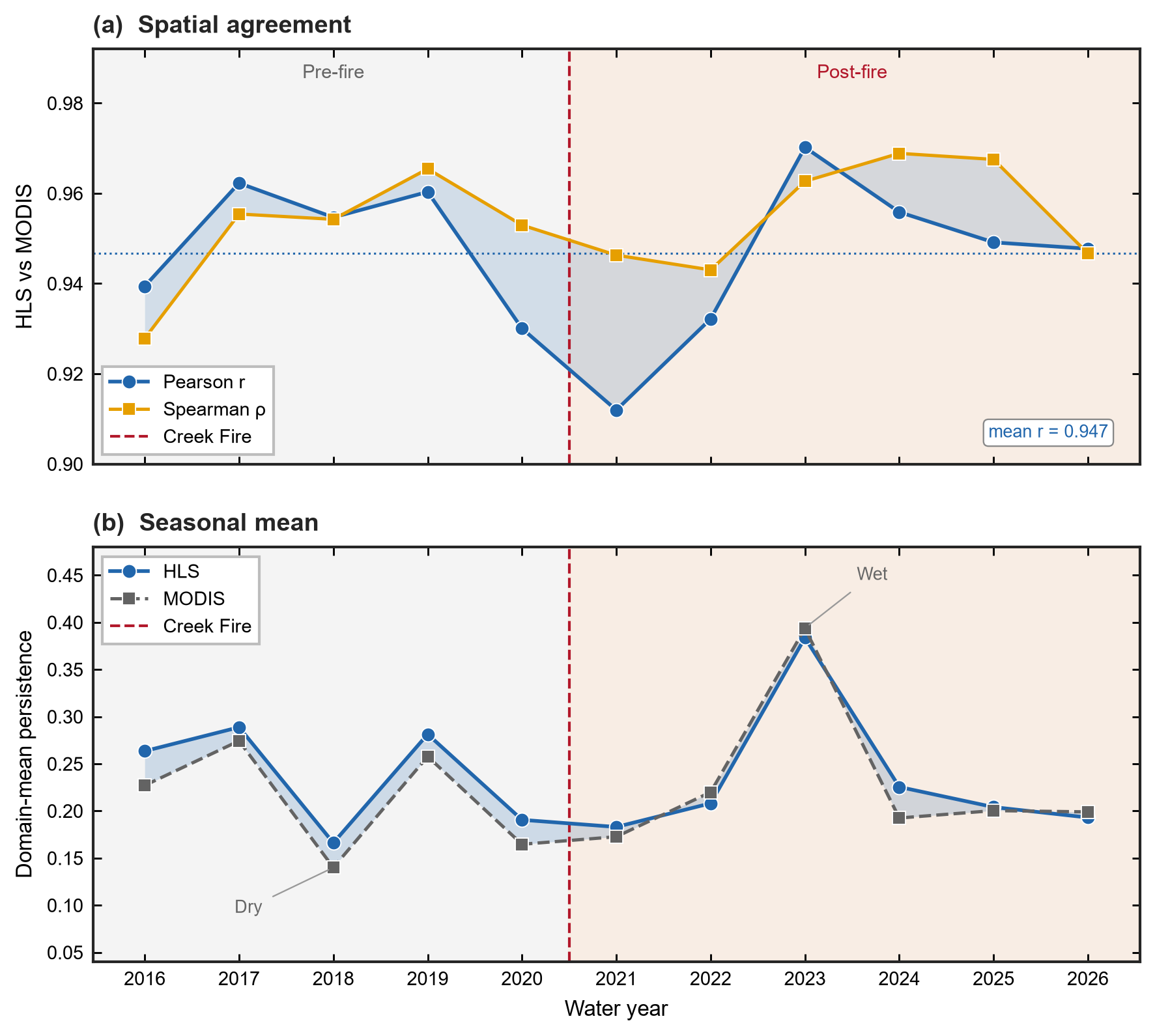}
\caption{HLS--MODIS consistency across the eleven-season record. Panel A shows annual spatial Pearson and Spearman correlations. Panel B compares domain-mean seasonal persistence. The dashed vertical line separates pre-fire and post-fire periods. The highlighted low- and high-persistence seasons illustrate shared interannual variation; the correlations are measures of agreement, not ground-reference classification accuracy.}
\label{fig:modis}
\end{figure*}

Annual spatial agreement between HLS and MODIS was high throughout the record. Pearson correlations ranged from 0.912 to 0.970, with a mean of 0.947; Spearman correlations ranged from 0.928 to 0.969 (Table~\ref{tab:modis}). The products also captured the same prominent annual contrasts, including the low-persistence WY2018 season and the high-persistence WY2023 season (Figure~\ref{fig:modis}).

\begin{table*}[!t]
\centering
\caption{Agreement between HLS and MODIS seasonal snow persistence.}
\label{tab:modis}
\footnotesize
\begin{tabularx}{\textwidth}{@{}Xl@{}}
\toprule
Quantity & Value \\
\midrule
Comparison period & WY2016--WY2026; 11 seasons \\
Domain-wide comparison cells per year & Approximately 66,300--66,600 \\
Annual Pearson correlation & 0.912--0.970 \\
Mean annual Pearson correlation & 0.947 \\
Annual Spearman correlation & 0.928--0.969 \\
Lowest HLS domain mean & 0.166, WY2018 \\
Highest HLS domain mean & 0.384, WY2023 \\
\bottomrule
\end{tabularx}\\[2pt]
\parbox{\textwidth}{\footnotesize These summaries refer to the domain-wide cross-sensor comparison, rather than the matched treatment--control sample.}
\end{table*}

The minimum and maximum HLS domain means were 0.166 in WY2018 and 0.384 in WY2023. HLS persistence was generally higher than MODIS persistence, commonly by approximately 0.01--0.04, although the direction was not uniform across all years. Differences in spatial support, observation timing, and snow detection provide plausible explanations for these offsets. Agreement between the products supports consistency of the broad snow patterns but does not establish which product more accurately represents snow beneath forest canopy.

\subsection{Predictive skill depends strongly on the target}\label{results-skill}
\begin{figure*}[!t]
\centering
\includegraphics[width=\textwidth]{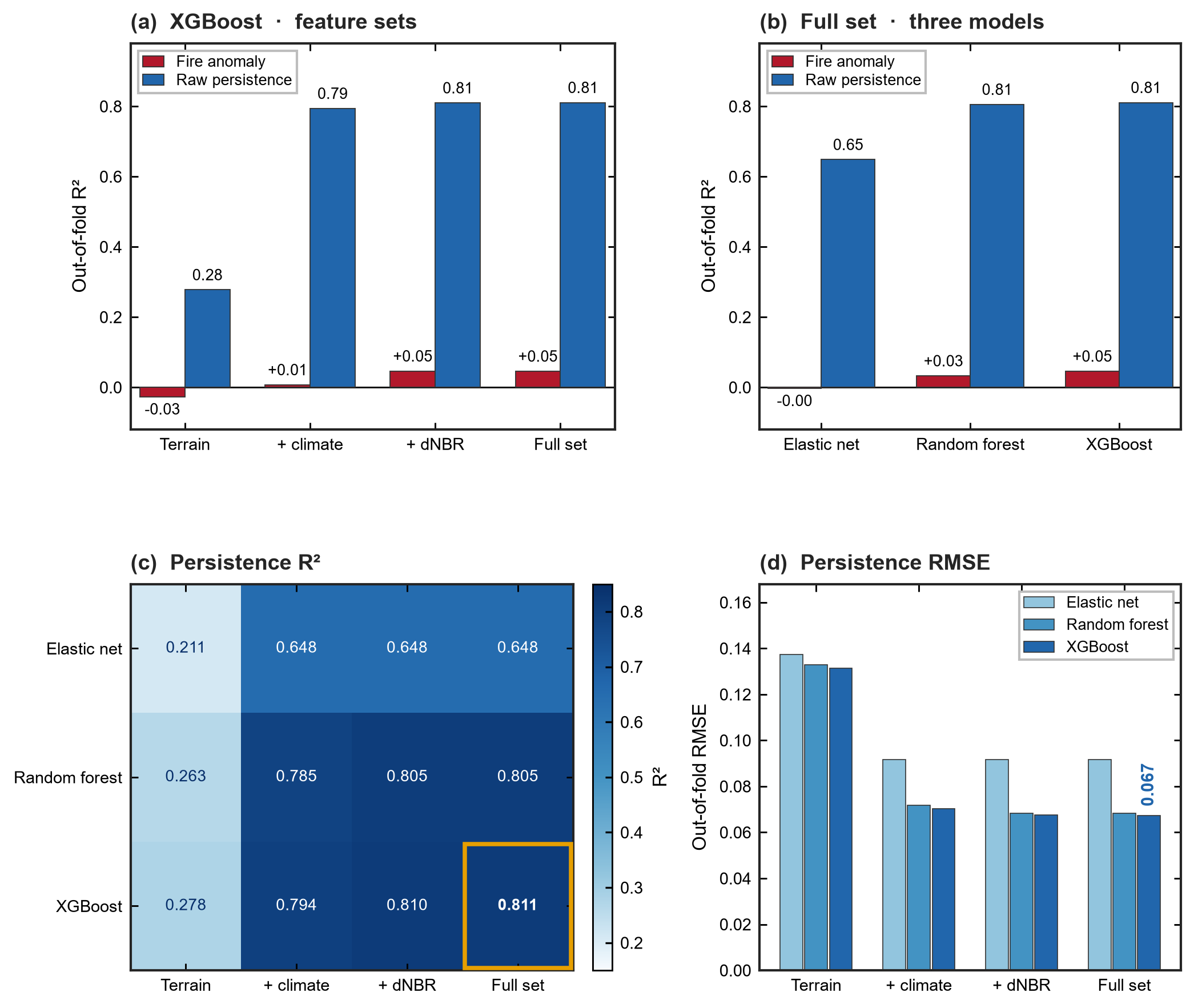}
\caption{Spatially evaluated prediction of raw persistence and the fire-adjusted anomaly. Panel A compares XGBoost feature configurations for the two targets. Panel B compares the three models using the full configuration. Panels C and D summarize raw-persistence $\Rsq$ and RMSE, respectively. Scores are out-of-fold estimates under 5~km spatial grouping. The ``+ dNBR'' configuration also includes years since fire. Plot labels are rounded; Table~\ref{tab:cv} reports three-decimal values.}
\label{fig:skill}
\end{figure*}

Raw post-fire persistence was substantially more predictable than the fire-adjusted anomaly. For XGBoost, terrain alone produced a persistence $\Rsq$ of 0.278. Adding meteorological information increased $\Rsq$ to 0.794, and the full configuration reached 0.811, with RMSE 0.067 and MAE 0.050 (Table~\ref{tab:cv}; Figure~\ref{fig:skill}). Random Forest achieved a similar full-configuration $\Rsq$ of 0.805, while Elastic Net reached 0.648.

\begin{table*}[!t]
\centering
\caption{Out-of-fold performance under five-fold, 5~km spatial GroupKFold.}
\label{tab:cv}
\footnotesize
\resizebox{\textwidth}{!}{%
\begin{tabular}{llccc}
\toprule
Target & Feature configuration & Elastic Net $\Rsq$ (RMSE) & Random Forest $\Rsq$ (RMSE) & XGBoost $\Rsq$ (RMSE) \\
\midrule
Fire-adjusted anomaly & A: terrain & $-0.003$ (0.083) & $-0.076$ (0.086) & $-0.027$ (0.084) \\
Fire-adjusted anomaly & B: $+$ meteorology & $-0.003$ (0.083) & $-0.017$ (0.083) & $+0.007$ (0.082) \\
Fire-adjusted anomaly & C: $+$ fire descriptors & $-0.003$ (0.083) & $+0.032$ (0.081) & $+0.046$ (0.081) \\
Fire-adjusted anomaly & D: full configuration & $-0.003$ (0.083) & $+0.032$ (0.081) & $+0.046$ (0.081) \\
Raw persistence & A: terrain & 0.211 (0.137) & 0.263 (0.133) & 0.278 (0.131) \\
Raw persistence & B: $+$ meteorology & 0.648 (0.092) & 0.785 (0.072) & 0.794 (0.070) \\
Raw persistence & C: $+$ fire descriptors & 0.648 (0.092) & 0.805 (0.068) & 0.810 (0.067) \\
Raw persistence & D: full configuration & 0.648 (0.092) & 0.805 (0.068) & 0.811 (0.067) \\
\bottomrule
\end{tabular}}\\[2pt]
\parbox{\textwidth}{\footnotesize $n=22{,}665$ cell-years. Configuration A contains elevation, slope, and northness. B adds domain-mean meteorology. C adds dNBR and years since fire. D denotes the full reported configuration. RMSE is expressed in persistence-fraction units.}
\end{table*}

The adjusted anomaly behaved differently. XGBoost $\Rsq$ was $-0.027$ with terrain alone and $+0.007$ after adding meteorology. The configuration containing dNBR and years since fire increased anomaly $\Rsq$ to $+0.046$, with no additional improvement in the reported full configuration. Random Forest reached $+0.032$, whereas Elastic Net remained at $-0.003$. Thus, the available predictors captured a small component of the adjusted spatial response, despite explaining much of the variation in raw persistence.

The improvement associated with feature set C concerns the combined addition of dNBR and years since fire; it cannot be assigned exclusively to dNBR. Similarly, the high raw-persistence score describes prediction within the observed post-fire years. Because domain-mean climate information was shared across spatial blocks, the models could use meteorological conditions already represented in the training data to distinguish those years.

\subsection{Terrain and meteorology dominate fitted-model attribution}\label{results-shap}
\begin{figure*}[!t]
\centering
\includegraphics[width=\textwidth]{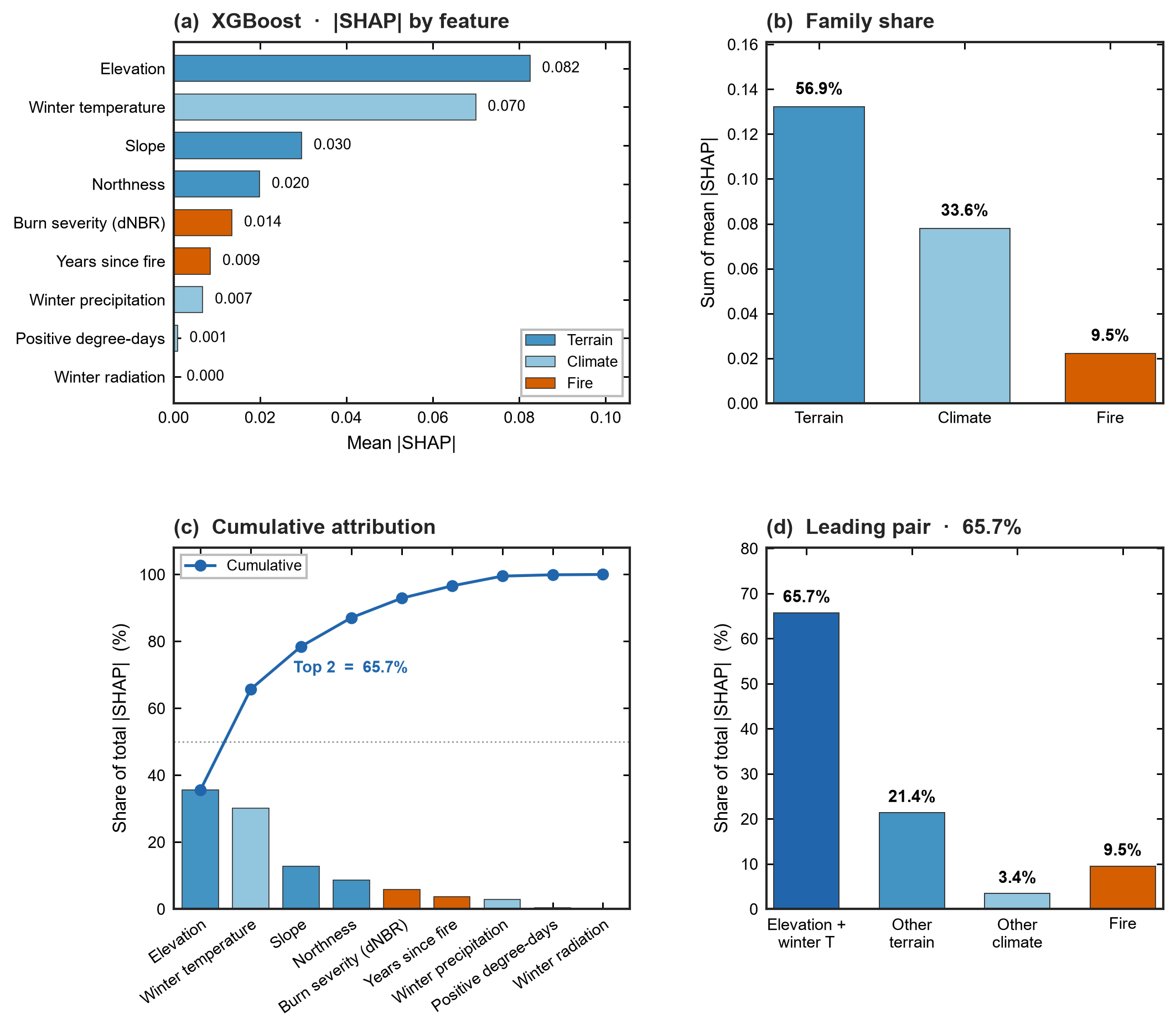}
\caption{SHAP attribution for the XGBoost raw-persistence model. Panel A ranks mean absolute SHAP values. Panel B aggregates attribution magnitude into terrain, climate, and fire families. Panel C shows cumulative attribution, and panel D separates elevation plus temperature from the remaining predictors. Attribution was calculated for 3,000 model rows. Family shares were calculated from the underlying values rather than the rounded labels displayed in panel A.}
\label{fig:shap}
\end{figure*}

The XGBoost persistence model assigned the greatest mean absolute SHAP values to elevation, 0.082, and temperature, 0.070. Slope and northness followed at 0.030 and 0.020, respectively. Burn severity and years since fire had mean absolute values of 0.014 and 0.009 (Figure~\ref{fig:shap}A).

Terrain accounted for 56.9\% of total attribution magnitude, climate for 33.6\%, and fire descriptors for 9.5\% (Figure~\ref{fig:shap}B). Elevation and temperature together accounted for 65.7\% (Figure~\ref{fig:shap}C,D). These shares describe how the fitted model distributed its predictions among the available predictors. They do not imply that terrain caused 56.9\% of snow variation or that fire caused only 9.5\% of the observed environmental response.

The attribution pattern is consistent with the model comparison: terrain and annual meteorological conditions explain much of the predictable raw-persistence structure, while the matched BACI analysis identifies a more specific response in the highest dNBR class.

\section{Discussion}\label{discussion}

\subsection{A severity-associated response within an unresolved landscape mean}\label{disc-severity}
The Creek Fire record supports two complementary findings. The landscape-average matched change was small relative to interannual variability, while the highest dNBR class showed increased observed snow persistence relative to comparable controls. These statements address different levels of aggregation. A weak perimeter-wide mean does not eliminate a response within a particular disturbance class, and a positive class response does not establish a uniform hydrological effect across the fire.

The high-class contrast of 0.026 represents a 2.6-percentage-point increase in the fraction of valid observations classified as snow. It is not a 2.6\% relative increase, an extension of the season by a specified number of days, or a measurement of additional snow water storage. The distinction follows directly from the observation-based definition of persistence.

The class pattern is physically plausible. Reduced interception after canopy disturbance can increase snow reaching the ground, while the balance between canopy shading and longwave radiation can change the duration of snow presence in ways that depend on winter conditions \citep{Varhola2010_01008009,Lundquist2013_rcr20504}. Field and modeling studies also demonstrate that snow response varies across severity and meteorological conditions \citep{Maxwell2019_ab5de8,Moeser2020_WR027071}.

At the same time, canopy opening can make existing ground snow more visible to an optical sensor. Validation against airborne lidar has shown that optical snow-mapping performance depends on forest cover and the spatial arrangement of snow and canopy \citep{Stillinger2023_75672023}. The observed positive contrast may therefore combine changes in ground snow occurrence with changes in detectability. Treating both pathways as part of the observation process makes the result more informative than interpreting increased optical persistence as an unqualified improvement in snow-water resources.

\subsection{Snow persistence complements, rather than replaces, snow-depth and disappearance metrics}\label{disc-metrics}
The Creek Fire findings do not conflict inherently with reports of earlier disappearance or reduced snow depth after wildfire. A seasonal observation-frequency metric summarizes snow presence across accumulation and ablation periods. Snow disappearance date emphasizes the final transition to snow-free conditions, whereas lidar depth quantifies the snowpack at specific acquisition times. Opposing changes within a season can affect these metrics differently.

\citet{Micheletty2014_46012014} provide a relevant Sierra Nevada comparison: their Moonlight Fire analysis reported increased fractional snow cover, demonstrating that a positive optical-cover response is not unprecedented. Conversely, \citet{Gleason2013_grl50896} showed that greater accumulation in a burned Oregon forest could coexist with earlier disappearance as increased radiative exposure accelerated melt. These examples emphasize the importance of distinguishing the quantity observed from the process inferred.

Recent work extends this reasoning across larger domains. \citet{Koshkin2025_vadt9866} examined how climate modifies post-fire snow disappearance, while \citet{Koshkin2026_34672026} used 114 airborne lidar acquisitions across nine Sierra Nevada basins to show contrasting accumulation- and ablation-season depth responses. The present analysis adds an eleven-season matched comparison of optical snow occurrence for one large fire. Its contribution is therefore complementary in temporal structure, observation type, and comparison design, rather than a claim that machine learning or Sierra Nevada fire--snow interactions have not previously been studied.

This measurement discipline also has broader relevance to GeoAI. Optical urban-canopy screening distinguishes visible crown candidates from a complete structural inventory, while farmland-boundary mapping distinguishes visible image boundaries from legal or operational parcel boundaries \citep{Narimani2026a_Farmland,Narimani2026b_Canopy}. In the same way, observed snow persistence should remain distinct from the underlying quantities of depth, water equivalent, and runoff. The model can only explain the target it has been trained to predict.

\subsection{High predictive accuracy and disturbance attribution answer different questions}\label{disc-prediction}
The contrast between the two modeling targets is a central result. XGBoost reproduced raw persistence with an out-of-fold $\Rsq$ of 0.811, yet reached only 0.046 for the matched, pre-fire-adjusted anomaly. The higher score reflects substantial predictable structure in elevation, terrain orientation, and the meteorological differences among the observed years. The lower score shows that these relationships do not transfer automatically to the residual disturbance-associated response.

This distinction is important for environmental applications of AI. A map can be accurate because it captures a strong background gradient, even when the disturbance signal of interest is small. Reporting both targets reveals what would be obscured by a single headline accuracy statistic. It also clarifies why a high-quality predictive model is not, on its own, an attribution model.

The evaluation design defines the domain of that accuracy. Spatial blocking withheld locations, but the six post-fire years remained represented across folds. The resulting score is appropriate for retrospective spatial prediction within the observed climatic record. It does not establish performance for a future winter, a different fire, or a different forest--snow regime. The same need to align evaluation with the intended use arises in other wildfire GeoAI applications, including spatially evaluated structure-loss modeling for the Palisades Fire \citep{FarajpoorNarimani2026_Palisades}, and is well established in ecological prediction \citep{Roberts2017_cog02881,Ploton2020_2018321y}.

SHAP strengthens interpretation when used within this predictive scope. The dominance of elevation and temperature shows which predictors the fitted persistence model used most strongly. It does not estimate the fraction of an ecological process caused by each variable. Moreover, predictors such as temperature and years since fire can share temporal information. Their attribution depends on the available feature set and should not be read as an independently identified physical mechanism.

The scientific value of the GeoAI component is thus not that a complex model necessarily outperforms every simpler alternative. Random Forest and XGBoost performed similarly for raw persistence, and both had modest anomaly skill. Rather, the modeling framework establishes which part of the post-fire record is readily predictable and which part remains weakly resolved by the available descriptors. This is consistent with the emphasis on interpretability, heterogeneous spatial settings, and meaningful evaluation in contemporary GeoAI research \citep{Mai2025_GeoAI}.

\subsection{Implications for forest monitoring and follow-up observation}\label{disc-monitoring}
The integrated framework offers a practical way to prioritize interpretation and subsequent measurement. Raw snow maps identify broad seasonal and topographic patterns. Matching reduces baseline differences between disturbed and reference locations. BACI contrasts then identify classes where the post-fire record diverges from the comparison. Machine learning evaluates whether the adjusted response can be mapped reliably from the available predictors.

For the Creek Fire, this sequence directs attention to the high-dNBR class without implying that every high-severity cell experienced the same response. The weak continuous severity correlation and substantial within-class variation favor targeted follow-up measurements rather than a uniform fire-wide response rule. Measurements of canopy structure, ground snow cover, depth, and seasonal radiation would be especially informative where optical persistence increased.

Several aspects of the present design define how broadly the findings should be applied. The 500~m analytical support averages local disturbance and snow conditions; almost half of the treatment cells fall in the low-change dNBR class. The severity thresholds are study-defined spectral categories, not calibrated estimates of tree mortality. HLS harmonization improves consistency between sensors but does not make acquisition frequency or clear-sky sampling constant through time \citep{Ju2025_HLS2}. Agreement with MODIS strengthens confidence in broad optical patterns, while leaving canopy visibility and shared optical sampling effects unresolved.

The matched controls also provide a regional comparison rather than a randomized experiment. Repeated use of some control cells and the single-fire temporal record remain relevant to uncertainty. Accordingly, the year-level intervals are more appropriate for the main inference than the narrower cell-level summaries alone. These considerations do not erase the observed class contrast; they identify the spatial, temporal, and observational conditions under which it is supported.

For forest and watershed monitoring, the main implication is methodological: interpret a disturbance response only after distinguishing it from background environmental predictability. The framework is useful because it can reveal both a positive class-level optical signal and a weakly predictable adjusted response within the same study. Neither result requires a claim that wildfire improves water supply or that a model has isolated all mechanisms governing post-fire snow.

\section{Conclusion}\label{conclusion}
An eleven-season multisource record of the 2020 Creek Fire shows a severity-associated increase in observed optical snow persistence without a resolved landscape-average change relative to matched controls. The highest dNBR class increased by 0.026, or 2.6 percentage points, while the overall contrast was $+0.0017$ with a year-level interval spanning zero.

Explainable GeoAI further separated environmental-state prediction from disturbance-response prediction. XGBoost achieved spatially evaluated $\Rsq=0.811$ for raw persistence but $\Rsq=0.046$ for the fire-adjusted anomaly. Elevation and temperature dominated fitted-model attribution. These results demonstrate why predictive accuracy should be interpreted alongside the target definition and comparison design. Combining multisource remote sensing, matched inference, and spatially evaluated machine learning provides a useful basis for monitoring post-fire forest--snow change without conflating visible snow, water storage, and causal disturbance effects.

\FloatBarrier

\section*{Data availability statement}\label{data-availability-statement}
The public source datasets used in this study are available through their respective HLS, MODIS, Sentinel-2, Copernicus DEM, ERA5-Land, and fire-perimeter repositories. Collection identifiers and processing information are provided in the Materials and methods. Derived persistence products, spectral-disturbance layers, matched-pair information, analytical tables, out-of-fold predictions, figure-source outputs, and supporting rasters are openly available on Zenodo at \url{https://zenodo.org/records/23024081}. Redistribution of source products follows the terms of the original providers.

\section*{Code availability statement}\label{code-availability-statement}
The code used for data processing, matching, statistical analysis, machine-learning evaluation, and figure preparation is available at \url{https://github.com/MohammadrezaNarimaniUCDavis/CreekFire_Snow_Persistence_GeoAI}. An archived replication release with environment specifications and reproducibility materials is included in the Zenodo record \url{https://zenodo.org/records/23024081}.

\section*{Author contributions}\label{author-contributions}
Parastoo Farajpoor: Conceptualization, Methodology, Validation, Visualization, Writing---original draft, Writing---review and editing.\\
Mohammadreza Narimani: Conceptualization, Data curation, Formal analysis, Methodology, Software, Validation, Visualization, Writing---original draft, Writing---review and editing, Project administration.

\section*{Funding}\label{funding}
The authors declare that no external financial support was received for the research, authorship, or publication of this article.

\section*{Conflict of interest}\label{conflict-of-interest}
The authors declare that the research was conducted in the absence of commercial or financial relationships that could be construed as a potential conflict of interest.

\section*{Ethics statement}\label{ethics-statement}
The study used geospatial and environmental observations and did not involve human participants, identifiable personal information, or experimental work with animals.

\section*{Generative AI statement}\label{generative-ai-statement}
During the preparation of this work, the authors used ChatGPT to improve grammatical accuracy, refine sentence structure, and enhance visualizations. All AI-generated revisions were thoroughly reviewed and edited by the authors to ensure relevance and accuracy.

\section*{Acknowledgments}\label{acknowledgments}
The authors acknowledge NASA, the National Snow and Ice Data Center, the European Space Agency, Copernicus, ECMWF, the MTBS program, Google Earth Engine, and the providers of the public geospatial resources used in this study. The study contains modified Copernicus Sentinel data. Terrain products were produced using Copernicus WorldDEM-30, \textcopyright\ DLR e.V.\ 2010--2014 and \textcopyright\ Airbus Defence and Space GmbH 2014--2018, provided under Copernicus by the European Union and ESA.

\section*{Abbreviations}\label{abbreviations}
BACI, before--after control--impact; dNBR, differenced Normalized Burn Ratio; HLS, Harmonized Landsat Sentinel-2; MAE, mean absolute error; MODIS, Moderate Resolution Imaging Spectroradiometer; MTBS, Monitoring Trends in Burn Severity; NBR, Normalized Burn Ratio; NDSI, Normalized Difference Snow Index; RMSE, root-mean-square error; SHAP, SHapley Additive exPlanations; XGBoost, extreme gradient boosting.

\bibliographystyle{IEEEtranN}
\bibliography{references}

\end{document}